\documentclass[9pt]{article}
\usepackage{spconf,amsmath,graphicx,amssymb}
\usepackage[subrefformat=parens]{subcaption}

\usepackage{hyperref}
\usepackage[detect-all]{siunitx}
\usepackage{mathtools}
\usepackage{comment}
\usepackage{multirow}
\usepackage{xspace}
\usepackage{booktabs}
\usepackage{mathtools}
\usepackage[textfont=it,tableposition=top]{caption}

\usepackage{blindtext}
\usepackage{todonotes}

\makeatother

\renewcommand{\v}[1]{\ensuremath{\mathbf{#1}}}

\title{Auxiliary Loss of Transformer with Residual Connection for End-to-End Speaker Diarization}

\name{Yechan Yu$^{1}$\textsuperscript{\textsection}   \quad
Dongkeon Park$^{2}$\textsuperscript{\textsection}   \quad
Hong Kook Kim$^{1,2}$
}

\address{
    $^{1}$School of Electrical Engineering and Computer Science,
    $^{2}$AI Graduate School,\\
    Gwangju Institute of Science and Technology, Gwangju 61005, Republic of Korea\\
    }

\email{yechan1202@gm.gist.ac.kr  \quad\quad  \{dongkeon,hongkook\}@gist.ac.kr}

\begin{document}
\ninept
\maketitle

%%%%%%%%%%%%%%%%%%%%%%%%%%%%%%%%%%%%%%%%%%%%%%%%%%%%%%%%%%%%%%%%%

\begin{abstract}
End-to-end neural diarization (EEND) with self-attention directly predicts speaker labels from inputs and enables the handling of overlapped speech. Although the EEND outperforms clustering-based speaker diarization (SD), it cannot be further improved by simply increasing the number of encoder blocks because the last encoder block is dominantly supervised compared with lower blocks. This paper proposes a new residual auxiliary EEND (RX-EEND) learning architecture for transformers to enforce the lower encoder blocks to learn more accurately. The auxiliary loss is applied to the output of each encoder block, including the last encoder block. The effect of auxiliary loss on the learning of the encoder blocks can be further increased by adding a residual connection between the encoder blocks of the EEND. Performance evaluation and ablation study reveal that the auxiliary loss in the proposed RX-EEND provides relative reductions in the diarization error rate (DER) by 50.3\% and 21.0\% on the simulated and CALLHOME (CH) datasets, respectively, compared with self-attentive EEND (SA-EEND). Furthermore, the residual connection used in RX-EEND further relatively reduces the DER by 8.1\% for CH dataset. 
\end{abstract}

%%%%%%%%%%%%%%%%%%%%%%%%%%%%%%%%%%%%%%%%%%%%%%%%%%%%%%%%%%%%%%%%%

\begingroup\renewcommand\thefootnote{\textsection}
\footnotetext{The first two authors contributed equally.}

\begingroup\renewcommand\thefootnote{}
\footnotetext{This research is supported by Ministry of Culture, Sports, and Tourism (MCST) and Korea Creative Content Agency (KOCCA) in the Culture Technology(CT) Research \& Development Program(R2020060002) 2022, and by the GIST-MIT Research Collaboration grant funded by the GIST in 2022.}

%%%%%%%%%%%%%%%%%%%%%%%%%%%%%%%%%%%%%%%%%%%%%%%%%%%%%%%%%%%%%%%%%

\section{Introduction}
\label{sec:intro}
% Speaker Diarization Definition 

Speaker diarization (SD) is a process for identifying “who spoke when” by dividing an audio recording into homogeneous segments using speaker labels \cite{tranter2003investigation, TrRe06, anguera2012speaker}. 
SD is essential to many speech-related applications with multi-speaker audio data, such as conversational multi-part speech recognition for business meetings or interviews and speaker-dependent video indexing \cite{reynolds05approaches,tranter04bn, amiwebsite}.
% Applications (ex. ASR, Enhancement)
% % Two Approach 1 (Clustering) = [Related Work]
% Speaker diarization work mainly proceeds in two directions. 
% One line is the clustering-based method, which is current state-of-the-art diarization systems that provide dependable performance in many challenges \cite{anguera2012speaker, DIHARD_data}. 
% The approach first segment a recording into short homogeneous blocks and compute speaker embeddings such as i-vectors \cite{Shum2013, Sell2014}, d-vectors \cite{Wan2018, Wang2018LSTM}, and x-vectors \cite{Snyder2018, Romero2017} are commonly used in speaker diarization tasks assuming that only one speaker in each block. 
% These embeddings are clustered to regroup segments by using clustering algorithms.
% However, most clustering-based diarization systems have several problems.
% They cannot handle overlapped speech since assuming embedding on that only one speaker in each block.
% Also, depend upon several modules (SAD, embedding extractor, overlap detection, etc.) that requires training independently, which complicates the system as a whole. 

In general, SD has been considered as a speaker clustering problem which assigns or classifies a speaker label to each speech segment. 
  A clustering-based SD system typically has a modular structure, comprising speech activity detection, a speaker embedding extractor, and speaker clustering \cite{Shum2013, Sell2014,senoussaoui2013study}.
For a given utterance, each segment is represented by a speaker embedding vector, such as i‑vectors \cite{Shum2013, Sell2014}, d-vectors \cite{Wan2018, Wang2018LSTM}, and x-vectors \cite{Snyder2018, Romero2017}. 
After assigning a speaker label to each segment, all segments with the same speaker label are grouped into a cluster.

Although such clustering-based SD systems have performed reliably in many recent SD challenges \cite{DIHARD_BUT, DIHARD_data, wang2021dkudukeecelenovo}, they mainly have two disadvantages.
First, they have difficulty in handling speech segments in which more than two speakers overlap because the speaker embedding for each segment can be represented by only one speaker.
Second, the performance of a clustering-based SD system is limited because all the modules are not jointly optimized.

End-to-end neural diarization (EEND) approaches have been proposed to overcome these disadvantages \cite{BLSTM-EEND, SA-EEND}. 
Unlike traditional clustering-based methods \cite{Snyder2018, Romero2017}, EEND methods consider SD as multi-label classification; therefore, speech activity detection and overlapped speech detection modules are unnecessary in the EEND framework
Consequently, the EEND method outperforms the clustering-based method for simulated and real datasets \cite{BLSTM-EEND, SA-EEND}.
Moreover, applying a self-attention mechanism to the EEND—self-attentive EEND (SA-EEND)—improves performance because self-attention could simultaneously accept global relation information over all frames \cite{SA-EEND}.

% % Advanced EEND
% Recently, several researchers have attempted to improve SA-EEND [x,y]. While SA-EEND provides well global attention over all frames, it is insufficient for SA-EENS to deal with local information regarding speaker changes over several adjacent frames. The time-dilated convolutional network (TDCN) was applied for a local embedding in SA-EEND [], which was a sequential architecture of modeling local and global information. On the other hand, a conformer-based EEND was proposed to improve the performance of SA-EEND, where the conformer was replaced with the transformer in SA-EEND to capture simultaneously local and global information [].
Recently, several researchers have attempted to improve SA-EEND \cite{maiti2021endtoend, CB-EEND}. 
Although SA-EEND provides adequate global attention over all frames, it is insufficient to deal with local information regarding speaker changes over several adjacent frames. 
The time-dilated convolutional network (TDCN), a sequential architecture of modeling local and global information, was applied to local embedding in SA-EEND \cite{maiti2021endtoend}. 
In contrast, a conformer-based EEND (CB-EEND) was proposed to improve the performance of SA-EEND, where the transformer was replaced with the conformer in SA-EEND to capture local and global information simultaneously \cite{CB-EEND}.

%%%%%%%%%%%%%%%%%%%%%%
% Problem Definition (What is challenge) 
% EEND Limitation

% Despite the achievements of EEND, this still remain a limitation.
% For other field using transformer block, many Encoder blocks are used to improve performance.
% However, we observe the performance is rather degraded when EEND use more than six Encoder blocks.
% In addition, only the last Encoder layer has a impact on diarization performance, and it seems as if the rest of the Encoder layers has not been learned effectively.
% While these approaches can improve performance, our preliminary experiment showed that SA-EEND could be further improved by enhancing the learning strategy of the transformer. The preliminary experiment was motivated by previous studies that increasing the number of encoder blocks in the transformer provided better performance for automatic speech recognition [] and natural language processing performance []. Accordingly, we increased the number of encoder blocks in the SA-EEND from four to six or more, but unfortunately SD performance degraded when the number of encoder blocks was greater than four. Furthermore, the encoder blocks near the input layer contributed little to the SD performance compared to those close to the output layer, which is discussed in Section 3. Therefore, the SD performance of SA-EEND is expected to be improved by enforcing the encoder blocks in the lower layer to learn more efficiently for a better contribution. 

Although these approaches can improve performance, our preliminary experiment demonstrated that SA-EEND could be further improved by enhancing the learning strategy of the transformer. 
The preliminary experiment was motivated by previous studies establishing that increasing the number of encoder blocks in the transformer improved performance for automatic speech recognition \cite{Speech-Transformer} and natural language processing \cite{BERT}.
Accordingly, we increased the number of encoder blocks in SA-EEND from four to or more.
Unfortunately, SD performance degraded when the number of encoder blocks was greater than four. 
Furthermore, the encoder blocks near the input layer contributed little to the performance compared with those close to the output layer, which will be explained in Sections \ref{sec:performance} and \ref{sec:analysis}.
Therefore, the performance of SA-EEND should be improved when enabling the encoder blocks in the lower layer to learn more accurately for a greater contribution. 

Based on this experiment, we propose a new residual auxiliary EEND (RX-EEND)-based learning architecture for transformers to enforce the lower encoder blocks to learn more accurately. 
There are two main contributions of this study: applying auxiliary loss and adding a residual connection.
% \begin{itemize}
% \item 
(1) When training the transformer in RX-EEND, the auxiliary loss is applied to the output of each encoder block, including the last encoder block.
The additional auxiliary loss should strongly supervise each encoder block, as observed in \cite{DETR}. 
Thus, the addition of encoder blocks improves performance, which will be discussed in Section \ref{sec:aux_loss}.
% \item 
(2) The effect of auxiliary loss on the learning of the encoder blocks can be further increased by adding a residual connection between the encoder blocks of the EEND because residual connections enable gradients to flow directly across the encoder blocks. 
Furthermore, using the residual block provides ensemble models as described in \cite{veit2016residual}.
Thus, the proposed RX-EEND should have a smaller generalization error than that of SA-EEND, which will be described in Section \ref{sec:residual}.
The rest of this paper is organized as follows. 
Section \ref{sec:SA-EEND} briefly reviews conventional SA-EEND, and Section \ref{sec:RX-EEND} proposes RX-EEND with auxiliary loss and residual connections. 
Next, Section \ref{sec:Experiment} evaluates the performance of the proposed RX-EEND on the simulated and CALLHOME (CH) datasets.
After that, an ablation study is conducted to demonstrate the individual contribution of auxiliary loss and residual connections each to the SD performance. In addition, the applicability of auxiliary loss and residual connections to conformer-EEND is discussed.
Finally, Section \ref{sec:conclusion} concludes the paper.

%%%%%%%%%%%%%%%%%%%%%%%%%%%%%%%%%%%%%%%%%%%%%%%%%%%%%%%%%%%%%%%%%
\section{SELF-ATTENTIVE EEND}
\label{sec:SA-EEND}
As described in the Introduction, the proposed RX-EEND is constructed based on the SA-EEND model \cite{SA-EEND} by adding a residual connection to each encoder block that is learned by its corresponding auxiliary loss. First, we briefly review conventional SA-EEND, as depicted in Fig.~\ref{fig:overview}(a). 

To extract input feature vectors, a 25-ms Hamming window with a hop size of 10-ms is applied to each utterance, and a 23-dimensional log-scaled mel-filterbank analysis is performed to each windowed speech frame.
Next, each series of 15 frames is concatenated into a single dimensional feature vector, $\v{x}_t\in \mathbb{R}^F$ with $F$~=~345, which is repeated with a stride of 10 frames, resulting in $X=[\v{x}_1,\cdots,\v{x}_T ]$, where $T$ is  the total number of feature vectors.

Next, $X$ is processed into a linear and layer normalization to obtain the input embedding vectors for the series of $P$ encoder blocks of EEND, $E^{0}=[{\v{e}_1^{0},\cdots,\v{e}_T^{0}}]$. Typically, $P$~=~4 in \cite{SA-EEND}. The $p$-th encoder block provides self-attentive features from the $(p-1)$-th embedding, as follows:
\begin{align}
    \v{e}^{0}_t &= \mathrm{Norm}(\mathrm{Linear}^{\textrm{F}}(\v{x}_t)) \in \mathbb{R}^{D}, \\
    %E^{0} &= Norm(Linear(X)) \in \mathbb{R}^{D}, \\
    E^{p} &= \mathrm{Encoder}_{p}^{\mathrm{D}}(E^{p-1}), \  (1 \le  p \le P). \label{eq:hidden}
\end{align}
% In the self-attentive EEND (SA-EEND) model \cite{SA-EEND}, speaker diarization is formulated as a multi-label classification problem. 
% The SA-EEND takes a $T$-length sequence which are sub-sampled by a factor of ten and  $F$-dimensional log-scaled Mel-filterbank based features , $X=[\v{x}_1, \dots, \v{x}_T], \ \v{x}_t \in R^F$, as an input, and  $\mathbf{W}_0 \in \mathbb{R}^{D \times F}$ and $\mathbf{b}_0 \in \mathbb{R}^D $ processes an input feature into an embedding $\v{e}_t^{(0)}\in\mathbb{R}^D$ at time index $t$.  % E_{0}=[\v{e}_1^{(0)}, \dots, \v{e}_T^{(0)}], \ 
% A $p$-th encoder block at time index $t$ denotes  $\mathrm{Encoder}^{(p)}_t(\cdot)$  which receives previous embedding vectors and outputs vector $\v{e}_t^{(p)}$ that has same input dimension. The total number of encoder blocks is $P$ and, input feature vector passes through all encoder blocks sequentially. After that, a linear transformation decoder $f:\mathbb{R}^D\to\mathbb{R}^S$ with an element-wise sigmoid function is applied to calculate posteriors 
% $\hat{\mathbf{y}}_t=[\hat{y}_{t,1},\dots,\hat{y}_{t,S}]\in(0,1)^S$ of 
% $S$ speakers at time  $t$.
% In the training phase, the SA-EEND is optimized using the PIT scheme \cite{yu2017permutation}, \ie , the loss is calculated between $\hat{\mathbf{y}}_t$ and the ground truth labels $\mathbf{y}_t=[y_{t,1},\dots,y_{t,S}]\in\{0,1\}^S$ as follows:
After passing all the $P$ encoder blocks, the last output vectors, $E^{P}$, are applied to a linear and sigmoid function to get the posteriors $\hat{\v{y}}_t=[\hat{y}_{t,1},\cdots,\hat{y}_{t,S}]$ of $S$ speakers at time $t$, as follows:
\begin{align}
    \hat{\v{y}}_{t} &= \mathrm{sigmoid}(\mathrm{Linear}^{\textrm{D}}(\v{e}_{t}^{P})) .
\end{align}
% \begin{align}
%     \v{e}^{(0)}_t &= \mathbf{W}_0\v{x}_t + \mathbf{b}_0 \in \mathbb{R}^{D}, \\
%     \v{e}^{(p)}_t &= \mathrm{Encoder}^{(p)}_t(\v{e}^{(p-1)}_1, \cdots, \v{e}^{(p-1)}_T) \  (1 \le  p \le P), \label{eq:hidden} \\
%     % \bigl[\mathbf{h}_{1,i},\ldots,\mathbf{h}_{T,i} \bigr] &=& \mathrm{Encoder}( \mathbf{X}_i ) \in \mathbb{R}^{D \times T}, \nonumber \\
%     \hat{\v{y}}_{t} &= \mathrm{sigmoid}(\mathrm{Linear}^{\textrm{D}}(\v{e}_{t}^{(P)})) \in (0,1).
% \end{align}
In the training phase, SA-EEND is optimized using the permutation invariant scheme \cite{BLSTM-EEND}.
That is, the loss is calculated between  $\hat{\v{y}}_t$  and the ground truth labels $\v{y}_t=[{y}_{t,1},\cdots,{y}_{t,S}] \in \{0,1\}^{S}$, thus the total loss function, $\mathcal{L}_{\textrm{d}}$ is defined as
%%% EQ - 1
\begin{align}
\mathcal{L}_{\textrm{d}} &= \frac{1}{TS}\min_{ \phi\in\mathrm{perm}(S)}\sum_{t=1}^{T}H\left(\mathbf{y}_t^\phi,\hat{\mathbf{y}}_t\right) \label{eq:diarization_loss}
\end{align}
%%%
% Same as \cite{SA-EEND}, the diarization loss is generated from diarization result and is formulated as:
where $\mathrm{perm}(S)$ is the set of all possible permutations of speakers, $\mathbf{y}_t^\phi \in \{0,1\}^S$ is the set of the permuted ground truth labels according to $\phi$.
In (\ref{eq:diarization_loss}), $H(\mathbf{y}_t,\hat{\mathbf{y}}_t)$ is the binary cross-entropy defined as 
%%% EQ - 5
\begin{align}
H\left(\mathbf{y}_t,\hat{\mathbf{y}}_t\right)=\sum_s{-y_{t,s}\log{\hat{y}_{t,s}}-\left(1-y_{t,s}\right)\log{\left(1-\hat{y}_{t,s}\right)}}. \label{eq:cross_entropy}
\end{align}
%%%
% where $\mathrm{perm}(1,\dots,S)$ is the set of all the possible permutation of speakers, $\mathbf{y}_t^\phi \in \{0,1\}^S$ is the permuted labels at $t$, and $H(\mathbf{y}_t,\hat{\mathbf{y}}_t)$ is the binary cross entropy determined as follows:

% \subsection{Conformer-based EEND}

% Recently, Conformer was proposed \cite{Conformer} as an ASR encoder architecture that combines local and global dependency modeling.
% Speaker diarization also depends both on local speaker changes and long-term speaker characteristics \cite{CB-EEND}.
% Each Conformer block, in turn, is composed of four stacked modules, including the first feed-forward network (FFN) module, a multi-head self-attention (MHSA) module, a convolution module, and the second FFN module. 
% The split feed-forward modules are known as sandwich structures inspired by \cite{lu2019understanding}. 
% Given the input $E_i$ to the $i$-th Conformer block, the output $E_{i+1}$ is computed as follows:
% % Encoder block using conformer
% \begin{equation} 
% \begin{aligned}
%  &\tilde{E}_i =E_i+\frac{1}{2}\text{FFN}(E_i)\\
%  &\tilde{E}_i'  = \tilde{E}_i + \text{MHSA}( \tilde{E}_i)\\
%  &\tilde{E}_i'' = \tilde{E}_i'+ \text{Convolution}(\tilde{E}_i')\\
%  &E_{i+1}     = \text{LayerNorm}(\tilde{E}_i''+\frac{1}{2}\text{FFN}(\tilde{E}_i''))
%  \end{aligned}
% \end{equation}

\section{PROPOSED METHOD}
\label{sec:RX-EEND}
In the conventional SA-EEND, the single loss function in (\ref{eq:diarization_loss}) is back-propagated into multiple number of encoder blocks sequentially. 
According to our preliminary experiment, the last encoder block contributes the most.
Moreover, increasing the number of encoder blocks from four to or more degraded the performance.
Thus, the RX-EEND is proposed to overcome this phenomenon by incorporating a new learning strategy and by adding a residual connection to each encoder block. 
The following subsections present a more detailed explanation for the proposed RX-EEND by comparing it with SA-EEND.

\begin{figure}[tb!]
 \begin{center}
  \includegraphics[width=42.5mm]{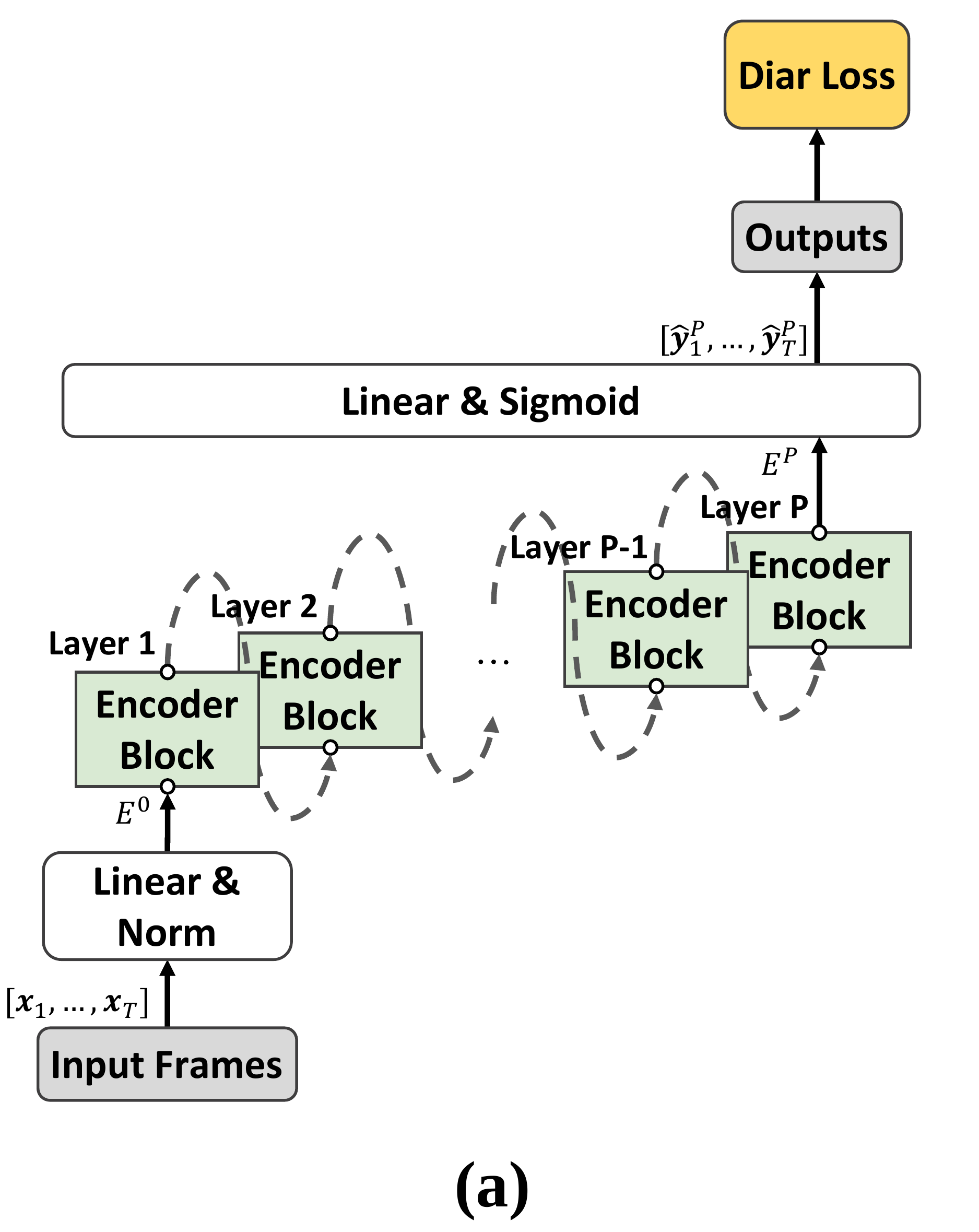}
  \includegraphics[width=42.5mm]{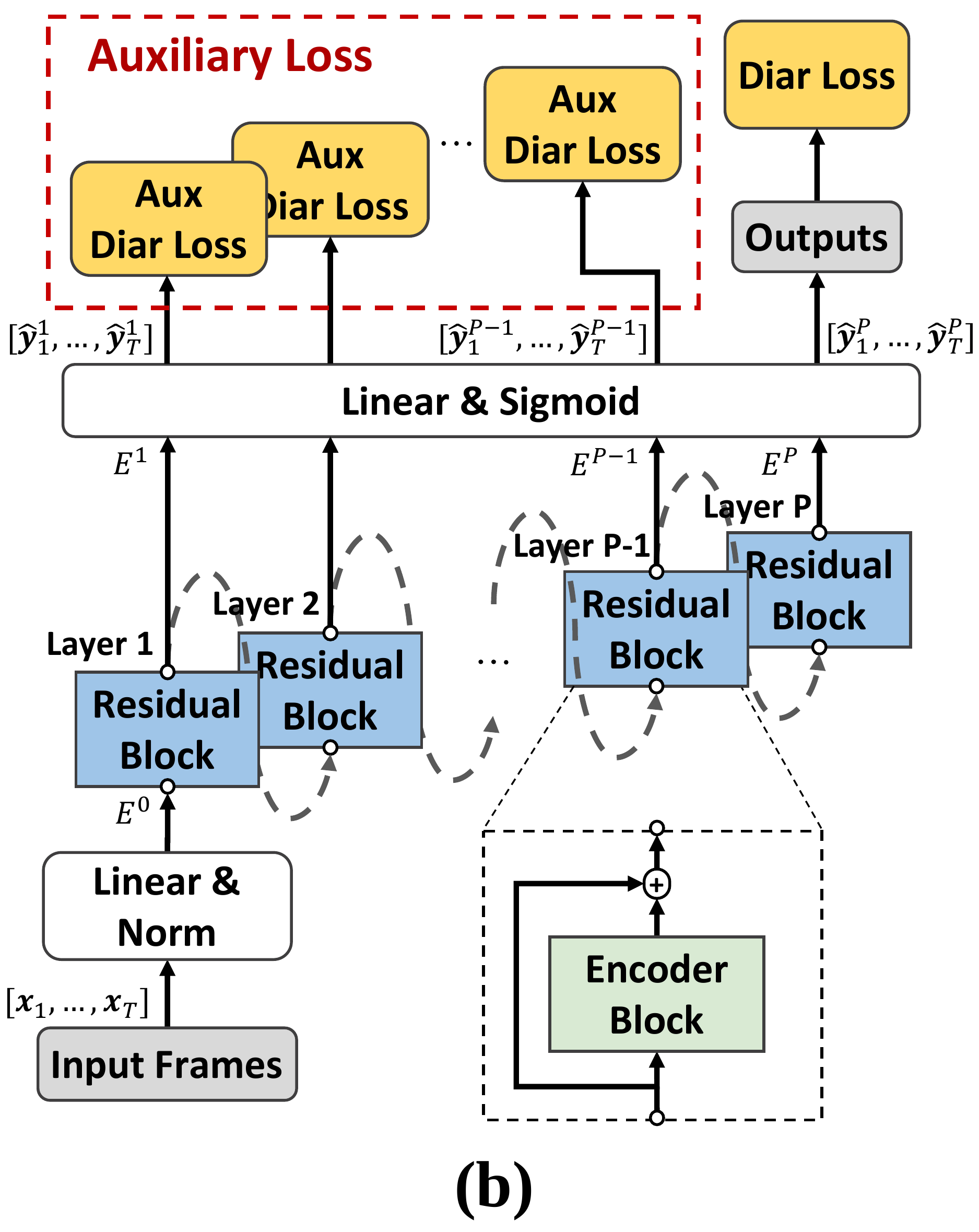}
 \end{center}
   \vspace{-6mm}
   \caption{
   Network architectures of (a) conventional SA-EEND \cite{SA-EEND} and (b) proposed RX-EEND.
}
 \label{fig:overview}
\end{figure}

\subsection{Auxiliary Loss}
\label{sec:aux_loss}
Inspired by the transformer-based object detection in \cite{DETR}, the first new approach for the proposed RX-EEND is to apply different auxiliary loss to each of encoder blocks. 
Accordingly, we apply a linear and sigmoid function to the outputs of each encoder block, whereas SA-EEND applies it only to the outputs of the last encoder block. 
As depicted in the upper part of Fig.~\ref{fig:overview}(b), the posteriors estimated from the $p$-th encoder embedding vectors, $\hat{\v{y}}_t^{p}$, are represented by 
\begin{align}
\hat{\v{y}}_{t}^{p} &= \mathrm{sigmoid}(\mathrm{Linear}^{\textrm{D}}(\v{e}_{t}^{p}))  \  (1 \le  p \le P-1).
\label{eq:aux_output}
\end{align}
Next, the loss function for training RX-EEND is defined as
\begin{align}
\mathcal{L}=\mathcal{L}_{\textrm{d}} + \lambda\mathcal{L}_{\textrm{aux}}
\label{eq:total_loss_with_aux}
\end{align}
where $\mathcal{L}_{\textrm{d}}$ is the same to (\ref{eq:diarization_loss}), and $\lambda$ is a hyperparameter to control the degree of effectiveness of the auxiliary loss against total loss.
The auxiliary loss can be defined in two ways, depending on how the permutation of speakers is applied. 
The first version of $\mathcal{L}_{\textrm{aux}}$ is defined as 
\begin{align}
\mathcal{L}_{\textrm{aux}}^{\textrm{shared}}&=\frac{1}{TS(P-1)}
\sum_{p=1}^{P-1}
\sum_{t=1}^{T}
H\left(\mathbf{y}_t^{\phi_{P}},\hat{\mathbf{y}}_{t}^{p}\right)
\label{eq:shared_aux_loss}
\end{align}
where $\phi_{P}$ is the same permutation obtained from (\ref{eq:diarization_loss}) and it is shared along all the lower encoder blocks. 
The second version of $\mathcal{L}_{\textrm{aux}}$ optimizes $\phi_{p}$ individually for each encoder block, as follows:
\begin{align}
\mathcal{L}_{\textrm{aux}}^{\textrm{indiv}}&=\frac{1}{TS(P-1)}
\sum_{p=1}^{P-1}
\min_{\phi_{p}\in\mathrm{perm}(S)}
\sum_{t=1}^{T}
H\left(\mathbf{y}_t^{\phi_{p}},\hat{\mathbf{y}}_{t}^{p}\right).
\label{eq:indivisual_aux_loss}
\end{align}
Based on the performance comparison for different auxiliary loss functions between (\ref{eq:shared_aux_loss}) and (\ref{eq:indivisual_aux_loss}), which will be discussed in Section \ref{sec:analysis}, $\mathcal{L}_{\textrm{aux}}^{\textrm{indiv}}$ in (\ref{eq:indivisual_aux_loss}) is selected as the auxiliary loss of the proposed RX-EEND.

\subsection{Residual Block}
\label{sec:residual}
As the second approach for the performance improvement of EEND, the encoder block in SA-EEND is modified by adding a residual connection to increase the convergence speed by directly propagating gradients from the $p$-th encoder block to the $(p-1)$-th encoder block. 
Hereafter, we refer to an encoder block with a residual connection as a residual block, as depicted in the zoomed box in Fig.~\ref{fig:overview}(b). Thus, the $p$-th residual block has the following function of 
%%%% Should describe residual block method
% As a means to both increase convergence speed and propagate signals more directly through the network, we deploy residual
% connections \cite{resnet} from one encoder block to the next. 
% Figure \ref{fig:overview} shows a diagram of the residual blocks.
% Changes from original EEND formula (Eq. \ref{eq:hidden}) are as follows:
\begin{align}
\v{e}^{p}_t &= \v{e}^{p-1}_t+\mathrm{Encoder}^{\textrm{D}}_p(\v{e}^{p-1}_1, \cdots, \v{e}^{p-1}_T) \  (1 \le  p \le P). \label{eq:residual}
\end{align}
The effectiveness of such residual blocks over encoder blocks will also be discussed in Sections \ref{sec:analysis} and \ref{sec:ablation}.

\begin{table}[t]
    \centering
    \scriptsize
    \caption{Distributions of simulated and real two speaker datasets for training and evaluating EENDs.}
    \vspace{-2mm}
    \resizebox{\linewidth}{!}{
    \begin{tabular}{ l l | c c }
    \toprule
    \multicolumn{2}{c}{Data Style} & \#Mixtures & Overlap ratio $\rho$ (\%)\\
    \midrule
    \textbf{Simulated Dataset}              &         &               &    \\
    \hspace{5mm}Sim2spk                     & {Train} & 100,000       &34.1\\
    \hspace{5mm}Sim2spk                     & {Test}  & 500/500/500   &34.4/27.3/19.6\\
    \midrule
    \textbf{Real Dataset}                   &         &               &    \\
    \hspace{5mm}CALLHOME \cite{callhome}    & {Train} & 155           & 14.0\\
    \hspace{5mm}CALLHOME \cite{callhome}    & {Test}  & 148           & 13.1\\
    \bottomrule
    \end{tabular}
    }
    \label{tbl:dataset}
    \vspace{-3mm}
\end{table}

%%%%%%%%%%%%%%%%%%%%%%%%%%%%%%%%%%%%%%%%%%%%%%

%%%%%%%%%%%%%%%%%%%%%%%%%%%%%%%%%%%%%%%%%%%%%%%%%%%%%%%%%%%%%%%%%

\section{Experiment}
\label{sec:Experiment}
\subsection{Datasets}
We trained and evaluated the proposed RX-EEND by preparing simulated and real datasets, as depicted in \autoref{tbl:dataset}.
For the simulated dataset, denoted as Sim2spk, we first obtained utterances in a speaker-wise manner from Switchboard-2 (Phases I, II, and III), Switchboard Cellular (Parts 1 and 2), and the NIST Speaker Recognition Evaluation (2004, 2005, 2006, and 2008). 
Then, we randomly chose 10 to 20 utterances for each speaker. 
Next, the utterances were convolved with one of the simulated room impulse responses (RIRs) used in \cite{rir} and added noise signal from the MUSAN corpus \cite{musan}.  
The noisy utterances from one speaker were mixed with those from the other speaker according to the overlap ratio. 
This procedure was repeated so that the total number of mixture files was up to 100,000. 
For the detailed procedure on generating the simulated dataset, see \cite{BLSTM-EEND}.
For the real dataset, we used the telephone conversation dataset CH \cite{callhome} (NIST SRE 2000; LDC2001S97, disk-8), which is the most widely used for SD studies. 
The CH dataset contained 500 sessions of multilingual telephonic speech.
Each session had two to six speakers, and two dominant speakers were in each conversation. 
The distribution of the CH dataset described in \autoref{tbl:dataset} is identical to that in \cite{SA-EEND}.
%%%%%%%%%%%%%%%%%%%%%%%%%%%%%%%%%%%%%%%%%%%%%%%%%%%%%%%%%%%%%%%%% % \begin{table}[t]
%     \centering
%     \caption{DERs (\%) on 2-speaker datasets.}
%     \label{tbl:results_2spk}
%     \resizebox{\linewidth}{!}{
%     \begin{tabular}{@{}lccccc@{}}
%         \toprule
%         &\multicolumn{3}{c}{Sim2spk}&\multicolumn{2}{c}{Real}\\\cmidrule(lr){2-4}\cmidrule(l){5-6}
%         Method& $\rho=\SI{34.4}{\percent}$&$\SI{27.3}{\percent}$&$\SI{19.6}{\percent}$&CH\\\midrule
%         %SA-EEND \cite{SA-EEND} & 4.56 &4.50 &3.85 &9.54\\
%         SA-EEND \cite{SA-EEND} &5.97 &5.65 &5.33 &10.72\\
%         %SA-EEND \cite{CB-EEND} &6.50 &N/A &N/A &10.52\\
%         TB-EEND \cite{CB-EEND} &3.54 &N/A &N/A &8.12\\
%         CB-EEND \cite{CB-EEND} &2.85 &N/A &N/A &9.70\\
%         RX-EEND &4.18 &3.93 &4.01 &9.17\\
%         SA-EEND deep &10.96 & & &12.15 \\
%         RX-EEND deep &3.13 & & &7.69 \\
%         RX-EEND Large &\textbf{2.74} &\textbf{2.45} &\textbf{2.76} &\textbf{7.37}\\
%         \bottomrule
%     \end{tabular}
%     }
% \end{table}

\begin{table}[t]
    \centering
    \scriptsize
    \caption{Performance comparison in DER(\%) between the proposed RX-EEND and SA-EEND.}
    \vspace{-2mm}
    \resizebox{\linewidth}{!}{
    \begin{tabular}{@{}lccccc@{}}
        \toprule
        &\multicolumn{3}{c}{Sim2spk}&\multicolumn{2}{c}{Real}\\\cmidrule(lr){2-4}\cmidrule(l){5-6}
        Method& $\rho$~=~34.4\%&$\rho$~=~27.3\%&$\rho$~=~19.6\%&CH\\
        \midrule
        SA-EEND \cite{SA-EEND}  &\ \ 5.97       &\ \ 5.65       &\ \ 5.33       &10.72\\
        %TB-EEND \cite{CB-EEND}  &3.54       &N/A        &N/A        &8.12\\
        %CB-EEND \cite{CB-EEND}  &2.85       &N/A        &N/A        &9.70\\
        RX-EEND                 &\ \ 4.18       &\ \ 3.93       &\ \ 4.01       &\ \ 9.17\\
        \midrule
        SA-EEND-deep            &10.33      &10.30      &\ \ 9.56       &12.62 \\
        RX-EEND-deep            &\ \ 3.13     &\ \ 2.84 &\ \ \bf 2.63   &\ \ 7.69 \\
        %RX-EEND-deep            &3.13       &2.84       &2.63       &7.69 \\
        \midrule
        SA-EEND-large           &\ \ 5.61       &\ \ 5.45       &\ \ 4.58       &10.15 \\
        RX-EEND-large           &\ \ \bf 2.74   &\ \ \bf 2.45   &\ \ 2.72       &\ \ \bf 7.37\\
        %RX-EEND-large           &2.74       &2.45       &2.72       &7.37\\
        \bottomrule
    \end{tabular}
    }
    \vspace{-3mm}
    \label{tbl:results_2spk}
\end{table}
%%%%%%%%%%%%%%%%%%%%%%%%%%%%%%%%%%%%%%%%%%%%%%%%%%%%%%%%%%%%%%%%%
\subsection{Experimental Setup}
\label{sec:Experimentalsettings}
To examine the performance over the baselines in SA-EEND, the proposed RX-EEND was first designed to have four residual blocks, 256 attention units with four heads ($H$~=~4, $D$~=~256 and $P$~=~4 ), and 1,024 internal units in a position-wise feed-forward layer, which was identical to those in SA-EEND  \cite{SA-EEND}. 
Then, the proposed RX-EEND and SA-EEND were trained using Sim2spk with $\rho$~=~34.1\%, as shown in the 1st row of \autoref{tbl:dataset}. 
On the other hand, the EENDs pretrained from Sim2spk dataset were finetuned for the real dataset.
As a performance metric, a diarization error rate (DER) \cite{nistrtwebsite} with a collar tolerance of \SI{0.25}{\s} between the predicted outputs and targets was calculated. 
In particular, the DER for Sim2spk was measured once every overlap ratio. 
Throughout the experiments, we set $\lambda$ in (\ref{eq:total_loss_with_aux}) to 1 by the exhaustive search, and the Adam optimizer \cite{adam} was used, where the learning rate schedule with warm-up steps of 100,000 was applied \cite{transformer}. And we set the number of training and adaptation epochs to be all 100 for simulated and real dataset, which setting was identical to that in \cite{SA-EEND}.

\subsection{Performance Comparison with SA-EEND}
\label{sec:performance}
\autoref{tbl:results_2spk} compares the DERs of the proposed RX-EEND with those of SA-EEND on the simulated and real datasets.
As shown in the first two rows of the table, RX-EEND outperformed SA-EEND for the two datasets. 
The lower DER for the real CH dataset implies that the residual connection of RX-EEND contributed adequately as a regularizer to reduce the generalization error.

Next, we investigated the effect of RX-EEND for deeper encoder or residual blocks by increasing the number of residual blocks of RX-EEND from four to eight. 
To this end, we set ($H$~=~4, $D$~=
~256 and $P$~=~8) in both SA-EEND and the proposed RX-EEND, which were denoted as SA-EEND-deep and RX-EEND-deep, respectively. 
The 3rd and 4th rows of \autoref{tbl:results_2spk} compare the DERs between SA-EEND-deep and RX-EEND-deep. 
As pointed out in the Introduction, the DER of SA-EEND-deep with eight encoder blocks was higher than that of SA-EEND with four encoder blocks. 
However, the DERs of RX-EEND-deep were much lower for the two datasets than those of RX-EEND.
Next, we observed the effectiveness of the auxiliary loss and residual connections for the different number of transformer heads and dimensions by changing $H$~=~4 and $D$~=~256 to $H$~=~8 and $D$~=~512 for both SA-EEND and RX-EEND, which were denoted as SA-EEND-large and RX-EEND-large in \autoref{tbl:results_2spk}. 
As shown in the last two rows of \autoref{tbl:results_2spk}, the DER reduction of SA-EEND-large was marginal compared with SA-EEND.
In contrast, RX-EEND-large reduced DERs for the real dataset and Sim2spk with overlap ratios of $\rho$~=~34.4\% and $\rho$~=~27.3\%.

% 우리는 TB-EEND와 CB-EEND 방법과 같은 추가적인 sub-sampling 방법과 encoder block 의 변경(conformer) 작업을 하지 않고, large model 만을 사용하여 기존 conventional EEND 보다 좋은 성능을 보였으며, 이후 이러한 작업을 추가 된다면 더 좋은 결과를 보일것으로 기대된다.

% 본 연구에서는 RX-EEND의 효율성을 비교하기 위해서, SA-EEND와 동일한 모델 hyperparameter 및 학습 전략을 사용하였습니다. TB-EEND는 기존 SA-EEND에 Convolutional Sub-samping 과 SpecAugmentation 이 추가되었으며, CB-EEND는 TB-EEND의 Encoder block을 Conformer 로 사용했습니다.

% To compare the efficiency of RX-EEND, we used the same model hyperparameter and learning strategy as SA-EEND. TB-EEND has been added Convolutional Sub-sampling and SpecAugmentation to existing SA-EEND, while CB-EEND has used TB-EEND's Encoder block as a Conformer.

% \begin{table*}[t]
% 	\caption{DER (\%) performance after each encoder layer on CALLHOME dataset.}
% 	\label{tab:tb_each_layer}
% 	\centering
% 	\footnotesize
% \begin{tabular}{lcccc cccc}
% 	\toprule
% 	 Method & Layer 1 & Layer 2 & Layer 3 & Layer 4 & Layer 5 & Layer 6 & Layer 7 & Layer 8 \\
% 	 \midrule
% 	 SA-EEND& 65.15 & 77.80& 72.97 & 74.76 & 77.19 & 40.63 & 53.68 & 10.15 \\
% 	 RX-EEND & 32.10 & 19.25 & 11.41 & 9.71 & 8.68 & 7.71 & 7.66 & \textbf{7.37} \\
% 	\bottomrule
% \end{tabular}
% \end{table*}

\begin{figure}
    \centering
    \begin{minipage}[c]{1\linewidth}
        \includegraphics[width=\linewidth]{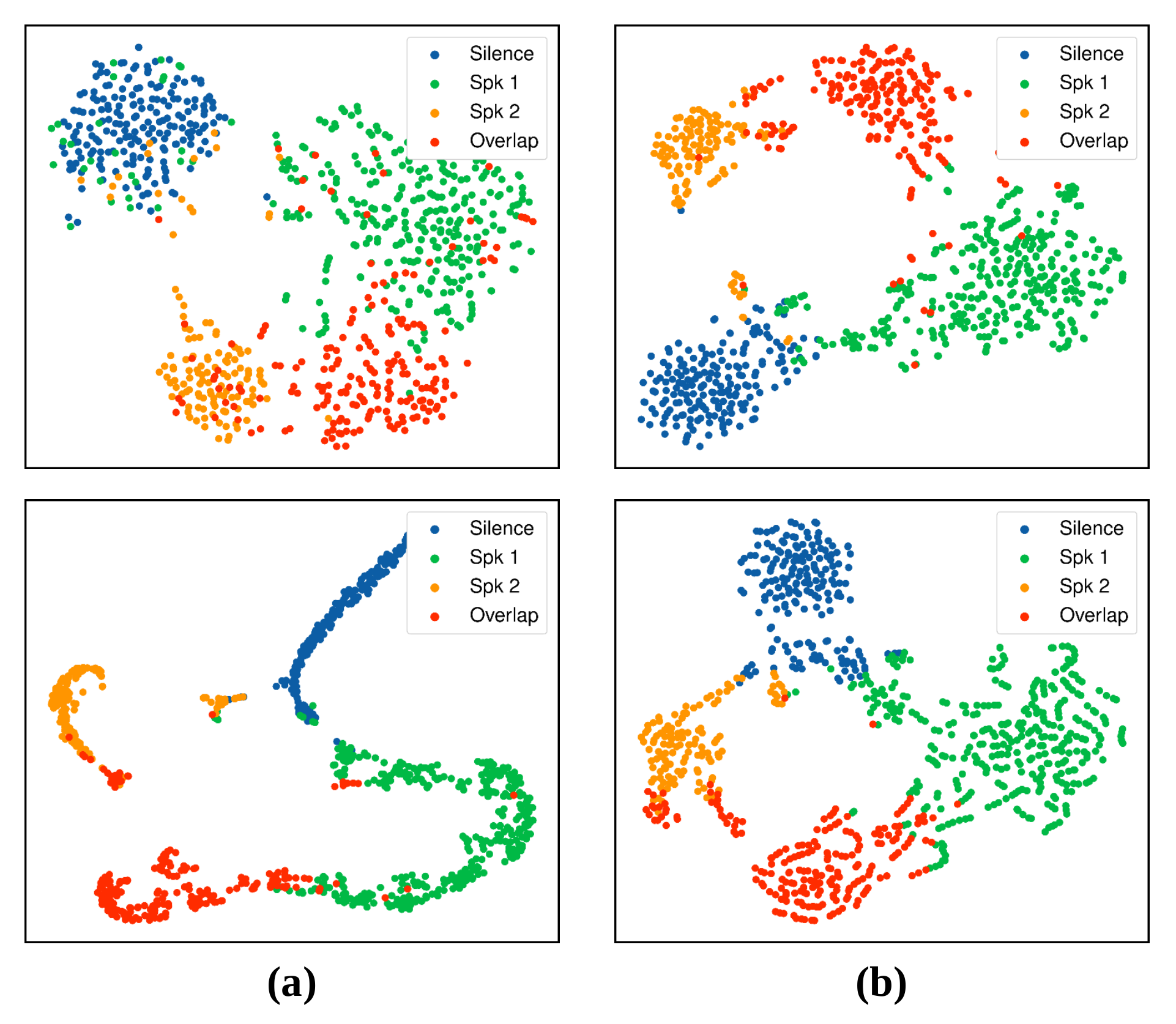}
    \vspace{-5mm}
    \end{minipage}
        \caption{
        T-SNE plots of embedding vectors extracted from the fourth and eighth blocks of (a) SA-EEND-large and (b) proposed RX-EEND-large applied to one mixture from Sim2spk with $\rho$~=~34.4\%.
      }
    \label{fig:visualization}
    \vspace{-1mm}
\end{figure}

\subsection{Analysis of the Contribution of Each Block}
\label{sec:analysis}
We examined the contribution of each encoder or residual block in SA-EEND or RX-EEND, respectively, by obtaining the embedding vectors at each block, $E^{p}$ for $p=1,\cdots,P$. 
Fig.~\ref{fig:visualization} illustrates the T-SNE plots of $E^4$ and $E^8$ for SA-EEND-large and RX-EEND-large, respectively, applied to one mixture from Sim2spk test data. % with $\rho$~=~34.4\%.
It was shown from the plots for $E^8$ that SA-EEND-large and RX-EEND-large distinctly separated silence and each speaker because $E^8$ was related to the actual output for the two models. 
However, SA-EEND-large failed to separate silence and speakers when $E^4$ was used. 
In contrast, the proposed RX-EEND-large could reasonably separate silence and speaker even with the lower block, $E^4$. 

Next, we predicted $\hat{\mathbf{y}}_t$ in (\ref{eq:aux_output}) using Sim2spk test data  with $\rho$~=~34.4\% for SA-EEND-large and RX-EEND-large, respectively. 
\autoref{tab:tb_each_layer} compares the DERs for each block of SA-EEND-large and RX-EEND-large.
The DER tended to decrease as the number of blocks increased for both models. 
However, except for the last block, the DERs at all the lower blocks for SA-EEND-large were extremely high, so that the performance in SA-EEND seemed to be highly dependent on the last block. 
Instead, the proposed RX-EEND-large could supervise each residual block to learn through the auxiliary loss and residual connection for a greater contribution. 
Consequently, these results support our motivation for proposing an auxiliary loss and a residual block.

\begin{table}[t]
	\centering
    \scriptsize
	\caption{Comparison of DERs (\%) of SA-EEND and RX-EEND using the embedding vectors at each block.}
    \vspace{-1mm}
    \resizebox{\linewidth}{!}{
	\setlength{\tabcolsep}{1mm}
    \begin{tabular}{lcccccccc}
	\toprule
	 Block number & 1 & 2 & 3 & 4 & 5 & 6 & 7 & 8 \\
	 \midrule
	 SA-EEND-large &82.25 &79.04 &64.70 & 71.75 &60.02 &40.86 &45.00 &\ \ 5.61 \\
	 RX-EEND-large &27.27 &13.46 &\ \ 7.28  &\ \ 4.95  &\ \ 4.06  &\ \ 3.16  &\ \ 2.89  &\ \ \textbf{2.74} \\
	\bottomrule
    \end{tabular}
    }
	\label{tab:tb_each_layer}
	\vspace{-3mm}
\end{table}
% \begin{table}[t]
%     \centering
%     \caption{DERs (\%) on 3-speaker datasets.}
%     \label{tbl:results_3spk}
%     \resizebox{\linewidth}{!}{
%     \begin{tabular}{@{}lcccc@{}}
%         \toprule
%         &\multicolumn{3}{c}{Sim3spk}&Real\\\cmidrule(lr){2-4}\cmidrule(l){5-5}
%         Method&$\rho=\SI{34.7}{\percent}$&$\SI{27.4}{\percent}$&$\SI{19.1}{\percent}$&CH\\\midrule
%         SA-EEND &8.69&7.64&6.92&14.00\\
%         SA-EEND + EDA (Shuffled) &8.38&7.06&6.21&13.92\\
%         RX-EEND &\textbf{5.16}&\textbf{-}&\textbf{-}&\textbf{12.86}\\
%         \bottomrule
%     \end{tabular}
%     }
% \end{table}

\subsection{Ablation Studies}
\label{sec:ablation}
Ablation studies were conducted to evaluate the effect of the auxiliary loss and residual blocks of the proposed RX-EEND on performance. 
\autoref{tbl:ablation_study} compares DERs on Sim2spk test data with $\rho$~=~34.4\% and CH dataset according to different variants of the proposed RX-EEND. 
Note here that SA-EEND-large and RX-EEND-large were used for these studies, and the 1st row of the table corresponds to the DERs for SA-EEND-large, which was identical to the DERs in \autoref{tbl:results_2spk}. 

First, we only replaced encoder blocks of SA-EEND-large with residual blocks. 
As shown in the 2nd row of the \autoref{tbl:ablation_study}, we achieved the lowered DERs by just adding residual connections to the encoder blocks, while the DER reduction was not significant.  
Next, we examined the performance difference between two different auxiliary loss types described in (\ref{eq:shared_aux_loss}) and (\ref{eq:indivisual_aux_loss}). 
We applied each auxiliary loss type to SA-EEND-large (i.e., we did not use residual connections to the encoder blocks of SA-EEND). 
As shown in the 3rd and 4th rows of the table, the two types reduced DERs, and the auxiliary type of (\ref{eq:indivisual_aux_loss}) was superior to that of (\ref{eq:shared_aux_loss}). 
In addition, the residual connections were more effective than auxiliary loss for the CH dataset, as depicted in the last row.

Finally, we examined whether the residual connections and auxiliary loss could be applied to other forms of EENDs.
Accordingly, CB-EEND \cite{CB-EEND} was chosen as a baseline of this study. 
For a fair comparison with SA-EEND and RX-EEND, SpecAugment and convolutional subsampling in CB-EEND were removed, which was referred to as Conformer-EEND in \autoref{tbl:comparsion_CB}.
We then constructed RX-conformer-EEND by adding residual connections to conformer blocks and training with auxiliary loss. 
As shown in \autoref{tbl:comparsion_CB}, the DERs of Conformer-EEND were higher than those of CB-EEND, implying that the subsampled features in CB-EEND could be dominant in DER reduction.
However, applying our proposed approaches to Conformer-EEND overcame this performance degradation; even the RX-conformer-EEND was significantly superior to CB-EEND in real dataset. 
Finally, this study revealed that the residual connections and auxiliary loss in RX-EEND could also improve performance for other forms of EENDs.

\begin{table}[t]
    \centering
    \scriptsize
    \caption{Ablation study of residual connections and auxiliary loss used for RX-EEND, measured in DER (\%). }
    \vspace{-2mm}
    \resizebox{\linewidth}{!}{
    \setlength{\tabcolsep}{3.5mm}
    \begin{tabular}{c c c c c}
        \toprule
        Residual &$\mathcal{L}_{\textrm{aux}}^{\textrm{shared}}$ &$\mathcal{L}_{\textrm{aux}}^{\textrm{indiv}}$ &Sim2spk & Real  \\
        \midrule
         & & & 5.61 & 10.15 \\
        \checkmark & & & 5.58  &10.00  \\
         & \checkmark & & 5.58 &\ \ 9.91  \\
         &  & \checkmark &2.79 &\ \ 8.02  \\
        \checkmark & & \checkmark & \textbf{2.74}  &\ \ \textbf{7.37}  \\
        \bottomrule
    \end{tabular}
    }
    \label{tbl:ablation_study}
    \label{tbl:comparsion_CB}
    \vspace{0mm}
\end{table}

\begin{table}[t]
    \centering
    \scriptsize
    \caption{Comparison of DERs (\%) when residual connections and auxiliary loss were applied to conformer-based EEND.}
    \vspace{-1mm}
    \resizebox{\linewidth}{!}{
    \begin{tabular}{@{}lccccc@{}}
        \toprule
        &\multicolumn{3}{c}{Sim2spk}&\multicolumn{2}{c}{Real}\\\cmidrule(lr){2-4}\cmidrule(l){5-6}
        Method& $\rho$~=~34.4\%&$\rho$~=~27.3\%&$\rho$~=~19.6\%&CH\\
        \midrule
        CB-EEND \cite{CB-EEND}  &2.85       &N/A        &N/A        &9.70\\
        Conformer-EEND          &3.90       &3.67       &3.89       &9.74\\
        RX-Conformer-EEND       &\bf2.79    &\bf2.54    &\bf2.34    &\bf8.31\\
        \bottomrule
    \end{tabular}
    }
    \label{tbl:comparsion_CB}
    \vspace{-3mm}
\end{table}
% \begin{table}[t]
%     \centering
%     \caption{Ablation Study on 2-speaker datasets.}
%     \label{tbl:ablation_study}
%     \setlength{\tabcolsep}{3mm}
%     \begin{tabular}{c c c c c}
%         \toprule
%         Residual &Auxiliary &Conformer &Sim2spk & CH  \\
%         \midrule
%          & & &4.56 & 9.54 \\
%         \checkmark & & &4.63  &9.52  \\
%          & \checkmark & &3.95 &8.66  \\
%         \checkmark & \checkmark& &3.75  &8.29  \\
%         \checkmark & \checkmark &\checkmark &2.67 &8.43 \\
%         \bottomrule
%     \end{tabular}
% \end{table}

%%%%%%%%%%%%%%%%%%%%%%%%%%%%%%%%%%%%%%%%%%%%%%%%%%%%%%%%%%%%%%%%%

\section{Conclusions}
\label{sec:conclusion}
In this paper, we proposed RX-EEND using auxiliary loss and residual connections to SA-EEND. 
The auxiliary loss supervised each encoder block individually to perform the same objective using residual connections in each encoder block to reduce the EEND generalization error. 
The proposed RX-EEND performance was evaluated on simulated and real datasets and compared with that of SA-EEND. 
The experimental results revealed that RX-EEND achieved significantly lower DERs than SA-EEND. 
Furthermore, based on the ablation studies, the residual connections or auxiliary loss contributed to reducing DERs and could be applied to other forms of EEND, such as CB-EEND. 
As future research, we will apply RX-EEND to a dataset with three or more speakers in future research.  

% \section{Acknowledgement}
% This research is supported by Ministry of Culture, Sports, and Tourism (MCST) and Korea Creative Content Agency (KOCCA) in the Culture Technology(CT) Research \& Development Program(R2020060002) 2020.

%%%%%%%%%%%%%%%%%%%%%%%%%%%%%%%%%%%%%%%%%%%%%%%%%%%%%%%%%%%%%%%%%

\vfill\pagebreak


\bibliographystyle{IEEEbib}
\bibliography{refs}

\end{document}